\documentclass[11pt,a4paper]{article} 

\usepackage{slashed}
\usepackage{hyperref}
\usepackage{color}
\usepackage{amsthm}
\usepackage{mathtools}
\usepackage[normalem]{ulem} 
\usepackage{graphicx}
\usepackage{amssymb,latexsym,cite}
\usepackage{amsmath}
\usepackage{amsfonts}
\usepackage{mathrsfs}
\usepackage{bbm}
\usepackage{bm}
\usepackage{cleveref}
\usepackage[T1]{fontenc}
\usepackage{multicol}
\usepackage{multirow}
\usepackage{float}

\newcommand{\CF}{\mathcal{F}}

\newcommand{\ii}{{\mathrm{i}}}
\newcommand{\frv}{\mathfrak{v}}
\newcommand{\Tr}{\mathrm{Tr}}
\newcommand{\nn}{\nonumber}

\def\RR{{\mathcal R}}

\def\nn{\nonumber}
\def\be{\begin{equation}}             \def\ee{\end{equation}}
\def\ba#1{\begin{array}{#1}}          \def\ea{\end{array}}
\def\bea{\begin{eqnarray} }           \def\eea{\end{eqnarray} }
\def\beann{\begin{eqnarray*} }        \def\eeann{\end{eqnarray*} }
\def\beal{\begin{eqalign}}            \def\eeal{\end{eqalign}}
             
\def\bsubeq{\begin{subequations}}     \def\esubeq{\end{subequations}}
\def\bitem{\begin{itemize}}           \def\eitem{\end{itemize}}
                    
\def\pa{\partial}

\def\a{\alpha}
\def\b{\beta}

\def\d{\delta}

\def\g{\gamma}

\def\i{\iota}

\def\k{\kappa}

\def\m{\mu}
\def\n{\nu}
\def\o{\omega}
\def\q{\theta}                  
\def\r{\rho}                    
\def\s{\sigma}                  
\def\t{\tau}

\def\x{\xi}

\newcommand{\sfR}{\mathsf{R}}

\def\L{\Lambda}
\def\O{\Omega}

\def\S{\Sigma}

\def\ii{\mathrm{i}} 
\def\diff{\mathrm{d}} 

\def\ll{\ell} 

\def\bpm{\begin{pmatrix}}
\def\epm{\end{pmatrix}}

\makeatletter
\def\@makechapterhead#1{%
  \vspace*{50\p@}%
  {\parindent \z@ \raggedright \normalfont
    \ifnum \c@secnumdepth >\m@ne
      \if@mainmatter
        \Huge\bfseries \thechapter\space%
      \fi
    \fi
    \interlinepenalty\@M
    \Huge \bfseries #1\par\nobreak
    \vskip 40\p@
  }}
\makeatother

\begin{document}

$\,$

\vspace{2.5cm}

\begin{center}

{{\LARGE{\bf Physical implications of four dimensional \\
braided noncommutative gravity}}}

\vspace*{1.5cm}

\baselineskip=14pt
		
		{\large\bf Milorad Be\v zani\'c}${}^{\,(a)\,,\,}$\footnote{Email: \ {\tt milorad.bezanic@ff.bg.ac.rs}} \ \ \ \ \ {\large\bf  Marija Dimitrijevi\'c \'Ciri\'c}${}^{\,(a)\,,\,}$\footnote{Email: \ {\tt
				dmarija@ipb.ac.rs}}  \\ {\large\bf Biljana Nikoli\'c}${}^{\,(a)\,,\,}$\footnote{Email: \ {\tt
				biljana@ipb.ac.rs}} \ \ \ \ \ {\large\bf Voja Radovanovi\'c}${}^{\,(a)\,,\,}$\footnote{Email: \ {\tt
				rvoja@ipb.ac.rs}} 
				\\[6mm]
		
		\noindent  ${}^{(a)}$ {\it Faculty of Physics, University of
			Belgrade}\\ {\it Studentski trg 12, 11000 Beograd, Serbia}
		\\[30mm]

\end{center}

\begin{abstract}
\noindent

The formulation of gravity theories on noncommutative (NC) spacetimes has been an active area of research for some time. Various models and methods have been proposed in the literature.  Even within the $\star$-product formalism, there are several distinct approaches to constructing NC deformations of General Relativity (GR) and Einstein–Cartan–Palatini (ECP) gravity. Some models are based on the NC $SO(1,3)_\star$ gauge symmetry, while others rely on NC deformations of the diffeomorphism symmetry of GR. In this paper, we investigate the connection between two NC gravity models based on twisted and braided NC diffeomorphism symmetry. Although these two models yield different NC gravity actions, we show that, in certain cases, they lead to the same  phenomenological consequences.  In particular, they share identical three-graviton vertex and exhibit some simple common vacuum solutions. This work represents a step towards understanding the diverse landscape of NC gravity models constructed within the $\star$-product framework.
\end{abstract}

\vspace*{2cm}



\newpage

{\baselineskip=14pt
\tableofcontents
}

\setcounter{footnote}{0}
\renewcommand{\thefootnote}{\arabic{footnote}}

\bigskip

\setcounter{page}{1}
\newcommand{\Section}[1]{\setcounter{equation}{0}\section{#1}}
\renewcommand{\theequation}{\arabic{section}.\arabic{equation}}

\section{Introduction}

Noncommutative (NC) gravity has been studied in the literature for more than 20 years. Various models have been constructed; for partial reviews, see \cite{Szabo:2006wx, Aschieri:2022kzo, DimitrijevicCiric:2022rmh}. In the $\star$-product approach, two main methods stand out:  $\star$-gauge theory of gravity, based on NC deformations of gauge symmetries such as the $SO(1,3)$ gauge group, and twisted/braided gravity, based on NC deformations of diffeomorphism symmetry.

In $\star$-gauge theory of gravity models, NC gravity is typically formulated in the first-order formalism, with the spin connection and vielbeins treated as independent fields. This method is based on $\star$-NC gauge theory with gauge groups such as $SO(1,3)\star$ or $SO(2,3)\star$, while the diffeomorphism symmetry is usually replaced by a twisted version. Since $\star$-gauge transformations close only in the universal enveloping algebra, one must either introduce new degrees of freedom \cite{Chamseddine_2004, deCesare:2018cjr} or employ the Seiberg–Witten map to express NC fields as functions of the corresponding commutative fields and the NC deformation parameter \cite{Aschieri:2022kzo, DimitrijevicCiric:2022rmh}. Expanding the NC gravity action in powers of the NC parameter yields the commutative gravity action plus corrections, with the first nonzero NC term being quadratic in the deformation parameter. The breaking of commutative diffeomorphism symmetry can be interpreted as fixing a specific reference frame, given by the Fermi normal coordinates of an observer. In that reference frame, noncommutativity is fixed to be of the Moyal type \cite{DimitrijevicCiric:2016uiq, DimitrijevicCiric:2016qio}.

More recently, the double copy formulation of $\star$-gauge theories with Moyal deformation was explored in \cite{Szabo:2023cmv}. The twisted version of colour-kinematics duality, compatible with the colour-kinematics mixing of NC $U(N)_\star$ gauge theory was developed. It was shown that the
result of applying the double copy relations in that context is commutative gravity. On the other hand, using the classical double copy prescription, the gravitational NC corrections arising from the double copy map of NC $SU(N)\star$ gauge theory were computed in \cite{Jonke:2025pkj}. As expected, these corrections are second order in the NC parameter and cubic in the powers of the curvature tensor.

A more geometrical and symmetry-based construction of an NC deformation of General Relativity (GR) is founded on a deformation of the (Hopf) algebra of diffeomorphisms. Specifically, using the Drinfel’d twist deformation method, the standard (commutative) diffeomorphism invariance of GR is replaced by a twisted version of this symmetry  \cite{Aschieri:2005yw}. The NC Einstein–Hilbert action, invariant under the twisted diffeomorphism symmetry, is then constructed.  Upon expansion in powers of the NC parameter, the first nontrivial NC correction  appears at quadratic order.This construction was further rigorously extended in \cite{Aschieri:2012ii, Aschieri:2016ody}, and NC gravity solutions were discussed in \cite{Ohl:2008tw, Ohl:2009pv}.

The braided NC deformation of Einstein–Cartan–Palatini (ECP) gravity, based on the $L_\infty$-algebra of ECP gravity \cite{Ciric:2020hfj}, was constructed in \cite{DimitrijevicCiric:2021jea}. As a gravity theory in the first-order formalism, the braided NC ECP gravity is based on braided diffeomorphism symmetry and braided $SO(1,3)$ gauge symmetry.  Unlike $SO(1,3)_\star$ gauge symmetry, the braided $SO(1,3)$ gauge symmetry closes in the standard $SO(1,3)$ algebra and does not require additional degrees of freedom or the Seiberg–Witten map. Ambiguities in the construction of the action are resolved by the structure of the cyclic braided $L_\infty$-algebra underlying the theory.

Given the multitude of NC gravity models based on the $\star$-product approach, it is natural to ask whether connections exist among them. A step in this direction was taken in \cite{Alvarez-Gaume:2006qrw}, where the next-to-leading order in the Seiberg–Witten limit of the gravitational action induced by bosonic string theory on a space-filling D-brane with a constant magnetic field was studied. It was shown that the induced terms for the three-graviton interaction vertex on the brane do not match  those derived from an action based on twisted diffeomorphism symmetry, as constructed in \cite{Aschieri:2005yw}.  While the low-energy limit of string theory reproduces the three-graviton vertex from the twisted diffeomorphism-based action, it also yields additional terms that cannot be obtained from the latter.

As a further step in comparing different NC gravity models, in this paper we analyze the twisted NC gravity constructed in \cite{Aschieri:2005yw} and the braided NC gravity of \cite{DimitrijevicCiric:2021jea}. Both are based on deformations of the Hopf algebra of diffeomorphisms, yet the resulting NC gravity actions are structurally different. We compare the three-graviton vertices obtained from both models, as well as with the three-graviton vertex induced by the low-energy limit of bosonic string theory on a space-filling D-brane with a constant magnetic field.

The paper is organized as follows. In the next section, we review braided NC ECP gravity in four dimensions and rewrite the corresponding braided $L_\infty$-algebra in a more standard $\gamma$-matrix basis, instead of the canonical basis used in \cite{Ciric:2020hfj}. The action in terms of $\gamma$-matrices cannot be uniquely written in arbitrary spacetime dimensions; however, this notation is more suitable for coupling spinor matter fields to gravity. In Section 3, we discuss the low-energy expansion, retaining only the first nontrivial NC correction to the commutative theory. As expected, this correction is quadratic in the NC deformation parameter. The commutative Minkowski spacetime and spatially flat de Sitter spacetime continue to be solutions of the braided NC gravity theory. As another application of the NC expansion, we compute the three-graviton vertex. Surprisingly, the result is the same as the previously obtained result for twisted Einstein gravity \cite{Alvarez-Gaume:2006qrw}. We conclude with comments and discussion in Section 5. Our conventions for four-dimensional $\gamma$-matrices and some detailed calculations are presented in the Appendices.

\section{Braided 4D ECP Gravity}

The braided $L_\infty$-algebra corresponding to the braided NC gravity in four spacetime dimensions was constructed in \cite{DimitrijevicCiric:2021jea}. In this section we briefly review this construction in a basis more suitable for coupling (spinor) matter fields to the gravitational filed.

\subsection{Commutative 4D ECP Gravity}

It is well known that the action functional for ECP gravity with a cosmological constant $\Lambda$ in four spacetime dimensions can be written as \begin{align}\label{eq:2.1}
S = \int \Tr \left(R\wedge e\wedge e\g_5 -\frac{\ii \L}{4!} e\wedge e\wedge e \wedge e \g_5 \right)\ .
\end{align}
The fundamental fields are the vierbein $e$ and the spin connection $\o$\footnote{Throughout this paper, we use Einstein summation convention: repeated upper and lower indices are implicitly summed over.}:
\begin{align}
e &= {e^a}_{\mu}\, {\rm d} x^\mu\, \g_a\ , \nn \\
\o &= \frac{1}{2}{\o^{ab}}_{\mu}\, {\rm d}x^\mu\, \S_{ab}\ , \label{eANDomega}
\end{align}
where
\begin{equation*}
 \S_{ab}=\frac{\ii}{4}[\gamma_a,\gamma_b]\ .
\end{equation*}
Some properties of the four-dimensional Dirac $\gamma$-matrices are listed in Appendix \ref{appendix:a}. The corresponding field strength tensors $R$ and $T$ represent the curvature and torsion, respectively, and are defined as\footnote{The commutator of Lie algebra-valued differential forms $\eta$ and $\zeta$, of degrees $|\eta|$ and $|\zeta|$, respectively, is given by
\begin{align*}
    [\eta,\zeta]=\eta\wedge \zeta - (-1)^{|\eta||\zeta|}\zeta \wedge \eta\ .
\end{align*}}
\begin{align}
R &= \frac{1}{4}{R^{ab}}_{\mu\nu}\, {\rm d}x^\mu\wedge {\rm d}x^\nu \, \S_{ab}= \mathrm{d}\o - \frac{\ii}{2}[\o,\o]\ , \nn\\
T &= \frac{1}{2}{T^a}_{\mu\nu}\, {\rm d}x^\mu \wedge {\rm d} x^\nu \, \g_{a} = \mathrm{d}e-\ii[\o,e]\ . \label{RandT}
\end{align}
Evaluating the traces of $\gamma$-matrices yields:
\begin{align}\label{eq:2.6}
S=\int \left( R^{ab}\wedge e^c \wedge e^d -\frac{\L}{6}\, e^a\wedge e^b \wedge e^c \wedge e^d\right)\varepsilon_{abcd}\  .
\end{align}
The vierbein $e^a$ and the spin connection $\omega^{ab}$ are real 1-forms, and due to the Hermitian properties of the $\gamma$-matrices, the following holds:
\begin{align}
e^\dagger=\g_0 e \g_0\ , \quad \o^\dagger=\g_0 \o \g_0\ ,
\end{align}
This can be used to show that the action \eqref{eq:2.1} is real.

By varying the action \eqref{eq:2.6} with respect to the vierbein and the spin connection, respectively, we obtain the following field equations:
\begin{align} 
0&= \varepsilon_{dabc}\left(e^a\wedge R^{bc}-\frac{ \Lambda}{3}\, e^a\wedge e^b\wedge e^c \right)\ , \label{eq:2.8} \\
0&= T^a\wedge e^b\ . \label{eq:2.9}
\end{align}
Equation \eqref{eq:2.9} represents the zero torsion condition and can be used to express the spin connection in terms of the vierbein and its derivatives. Inserting the resulting solution $\omega(e)$ into \eqref{eq:2.8}, and rewriting the equation in terms of the metric tensor $g_{\mu\nu} = \eta_{ab} e^a_{\ \mu} e^b_{\ \nu}$, reproduces the standard Einstein equations of General Relativity.

The action functional \eqref{eq:2.1} is invariant under both diffeomorphisms and $SO(1,3)$ gauge transformations:
\begin{align}
\d_\x e &= \mathrm{L}_\x e\ , \quad \d_\x \o=\mathrm{L}_\x \o\ , \label{eq:2.13} \\
\d_\r e &= \ii[\r,e]\ , \quad \d_\r \o=\mathrm{d}\r+\ii[\r,\o]\ , \label{eq:2.14}
\end{align}
with the parameters
\begin{align}
 \x = \x^\mu (x) \partial_\mu, \quad  \r=\frac{1}{2}\r^{ab} (x)\S_{ab}\ \nn
\end{align}
and the Lie derivative $\mathrm{L}_\x$ acting on a $p$-form $\tau$ as
\begin{equation}
\mathrm{L}_\xi \tau = \mathrm{d}\circ\iota_\xi \tau + \iota_\xi \circ \mathrm{d} \tau\ . \nn
\end{equation}
The Noether identities, which correspond to the invariance of the action \eqref{eq:2.1} under  $SO(1,3)$ gauge transformations and diffeomorphisms (\ref{eq:2.13}, \ref{eq:2.14}) relate the Euler-Lagrange derivatives off-shell. They can be derived starting from the Euler-Lagrange variation of the action:
\begin{equation}
\delta S = \int \Tr \left( \delta e\wedge \mathcal{F}_e \g_5  + \delta \omega\wedge \mathcal{F}_\omega \g_5 \right)
\end{equation}
and inserting the explicit transformation laws of the vierbein and the spin connection (\ref{eq:2.13}, \ref{eq:2.14}). This results in the diffeomorphism Noether identity:
\begin{align}
\mathrm{d}x^\m \otimes \Tr \left(((\i_\m \mathrm{d}e)\wedge \mathcal{F}_e +(\i_\mu \mathrm{d}\o)\wedge \mathcal{F}_\o - \i_\mu e (\mathrm{d}\mathcal{F}_e)-\i_\mu \o(\mathrm{d}\mathcal{F}_\o))\g_5\right) =0\ , \label{DiffNoetherId}
\end{align}
and the $SO(1,3)$ gauge Noether identity:
\begin{align}
-\mathrm{d}\mathcal{F}_\o +\ii e\wedge \mathcal{F}_e -\ii\mathcal{F}_e \wedge e+\ii[\o,\mathcal{F}_\o] =0 \ , \label{LLNoetherId}
\end{align}
where $\i_\m$ denotes contraction along the basis vector $\pa_\m$.

\vspace{5mm}
{\bf $L_\infty$-algebra of the 4D ECP gravity}
\vspace{3mm}

The complete data of classical ECP gravity are contained in the corresponding $L_\infty$-algebra. The graded vector space $V$ underlying this $L_\infty$-algebra is
\begin{align} \label{eq:2.18}
    V= V_0 \oplus V_1 \oplus V_2 \oplus V_3\ ,
\end{align}
where the following notation is used for the homogeneous elements\footnote{The degree of a homogeneous element $v\in V_k$ is $k$.}:
\begin{align} \label{eq:2.18_2}
    \begin{pmatrix}
        \xi\\
        \rho
    \end{pmatrix}\in V_0\ , \quad
    \begin{pmatrix}
        e\\
        \o
    \end{pmatrix}\in V_1\ , \quad
    \begin{pmatrix}
        E\\
        \O
    \end{pmatrix}\in V_2\ , \quad
    \begin{pmatrix}
        \mathcal{X}\\
        \mathcal{P}
    \end{pmatrix}\in V_3\ ,
\end{align}
and the Euler-Lagrange derivatives corresponding to the fields $e$ and $\o$ are denoted as $E$ and $\O$, respectively, while the Noether identities corresponding to the diffeomorphism and $SO(1,3)$ gauge symmetries are denoted as $\mathcal{X}$ and $\mathcal{P}$, respectively. The only nonvanishing $1$-brackets are defined by
\begin{align}
\begin{split}
    \ell_1 \begin{pmatrix}
        \xi\\
        \rho
    \end{pmatrix} =\begin{pmatrix}
        0\\
        \mathrm{d}\r
    \end{pmatrix}\in V_1\ ,\quad\quad \ll_1 \bpm
        E\\
        \O
    \epm =\bpm
        0\\
        -\mathrm{d}\O
    \epm\in V_3\ .
\end{split}
\end{align}
The nonvanishing $2$-brackets are defined by
\begin{align} 
    \begin{split} \label{eq:2.21}
        \ll_2\left(
    \begin{pmatrix}
    \xi_1\\
    \rho_1
    \end{pmatrix},
    \begin{pmatrix}
        \xi_2\\
        \rho_2
    \end{pmatrix}
    \right)&=
    \begin{pmatrix}
        [\xi_1,\xi_2]\\
        \mathrm{L}_{\xi_1}\rho_2-\mathrm{L}_{\xi_2}\rho_1+\ii[\rho_1,\rho_2]
    \end{pmatrix} \in V_0
    \ ,\\
     \ll_2\left(
     \begin{pmatrix}
         \xi\\
         \rho
     \end{pmatrix},
     \begin{pmatrix}
         e\\
         \omega
     \end{pmatrix}
     \right)&=
     \begin{pmatrix}
         \mathrm{L}_\xi e+\ii[\rho,e]\\
         \mathrm{L}_\xi \omega+\ii[\rho,\omega]
     \end{pmatrix}\in V_1
     \ ,\\
      \ll_2\left(
        \begin{pmatrix}
            \xi\\
            \rho
        \end{pmatrix},
        \begin{pmatrix}
            E\\
            \Omega
        \end{pmatrix}
    \right)&=
    \begin{pmatrix}
         \mathrm{L}_\xi E+\ii[\rho,E]\\
         \mathrm{L}_\xi \Omega+\ii[\rho,\Omega]
    \end{pmatrix}\in V_2 \ , \\
\ll_2 \left(
        \begin{pmatrix}
            \xi\\
            \rho
        \end{pmatrix},
        \begin{pmatrix}
            \mathcal{X}\\
            \mathcal{P}
        \end{pmatrix}
    \right)&=
    \begin{pmatrix}
            \mathrm{L}_\xi \mathcal{X} +  \mathrm{d}x^\mu \otimes \Tr ((\iota_\mu \mathrm{d} \rho)  \mathcal{P} \gamma_5)\\
            \mathrm{L}_\xi \mathcal{P}+\ii[\rho,\mathcal{P}]
        \end{pmatrix}\in V_3  \ ,\\
  &\nn\\
       \ll_2\left(
       \begin{pmatrix}
           e_1\\
           \omega_1
       \end{pmatrix},
       \begin{pmatrix}
           e_2\\
           \omega_2
       \end{pmatrix}
      \right)&=
      \begin{pmatrix}
          -(e_1\wedge \mathrm{d}\omega_2 +\mathrm{d}\omega_2\wedge e_1+_{(1\leftrightarrow 2)})\\
          -\mathrm{d}[e_1,e_2]
      \end{pmatrix}\in V_2
        \ ,\\
    \ll_2\left(\begin{pmatrix}
        E\\
        \Omega
    \end{pmatrix} ,
    \begin{pmatrix}
        e\\
        \omega
    \end{pmatrix}
    \right)&=\begin{pmatrix}
        \mathrm{d}x^\mu \otimes \Tr (((\iota_\mu \mathrm{d}e)\wedge E+(\iota_\mu\mathrm{d}\omega) \wedge \Omega -(\iota_\mu e)\mathrm{d}E-(\iota_\mu \omega)d\Omega)\gamma_5)\\
        \ii e\wedge E-\ii E\wedge e+\ii[\omega,\Omega]
    \end{pmatrix}\in V_3 \ .
    \end{split}
\end{align}
The only nonvanishing $3$-bracket is defined by
\begin{align}
\begin{split}
    \ll_3&\left(
    \begin{pmatrix}
        e_1\\
        \omega_1
    \end{pmatrix},
    \begin{pmatrix}
        e_2\\
        \omega_2
    \end{pmatrix},
    \begin{pmatrix}
        e_3\\
        \omega_3
    \end{pmatrix} \right)\\
    &=
    \bpm
        \ii\Big(
            e_1\wedge [\omega_2,\omega_3]+[\omega_2,\omega_3]\wedge e_1+\frac{\Lambda}{6} e_1\wedge[e_2,e_3]+_{(1\leftrightarrow 2)}+_{(1\leftrightarrow 3)}\Big)\\
        \ii\Big(\big[\o_1, [e_2, e_3]\big] + \big[\o_2, [e_1, e_3]\big] + \big[\o_3, [e_1, e_2]\big] \Big)
    \epm   \in V_2\ .
\end{split}
\end{align}

These brackets straightforwardly satisfy the corresponding homotopy relations; see Appendix \ref{appendix:b} for details.

A non-degenerate bilinear pairing
$\langle-,-\rangle :V_p \otimes V_q \rightarrow \mathbb{R}, \>\> p+q=3$ promotes the $L_\infty$-algebra to a cyclic $L_\infty$-algebra and enables the use of the variational principle. In our example, the pairing is defined as follows:
\begin{align} 
& \Big\langle \begin{pmatrix}
        e\\
        \omega
    \end{pmatrix},
    \begin{pmatrix}
        E\\
        \Omega
    \end{pmatrix}
    \Big\rangle = \ \int \Tr \big((e\wedge E + \omega \wedge \Omega)\gamma_5\big)\ , \label{eq:2.23}\\
& \Big \langle
    \begin{pmatrix}
        \xi\\
        \rho
    \end{pmatrix},
    \begin{pmatrix}
        \mathcal{X}\\
        \mathcal{P}
    \end{pmatrix}
    \Big \rangle
    = -\int \iota_\xi \mathcal{X}-\int \Tr  (\rho \mathcal{P} \gamma_5)\ . \nn
\end{align}
The ECP action functional (\ref{eq:2.1}) can be recovered as
\begin{align}
    S=-\frac{1}{3!}\left\langle \bpm
    e\\
    \o
    \epm,\ll_2\left(
       \begin{pmatrix}
           e\\
           \omega
       \end{pmatrix},
       \begin{pmatrix}
           e\\
           \omega
       \end{pmatrix}
      \right)\right\rangle
      -\frac{1}{4!}\left\langle \bpm
    e\\
    \o
    \epm,\ll_3\left(
       \begin{pmatrix}
           e\\
           \omega
       \end{pmatrix},
       \begin{pmatrix}
           e\\
           \omega
       \end{pmatrix},
       \bpm
            e\\
            \o
       \epm
      \right)\right\rangle\ .
\end{align}

\subsection{Braided $L_\infty$ algebra of 4D ECP gravity}

Starting from a suitable $L_\infty$-algebra and using Drinfel'd twist deformation techniques, one can construct a braided $L_\infty$-algebra in the sense of \cite{Ciric:2020eab, DimitrijevicCiric:2021jea}. In the Drinfel'd twist formalism, a deformation is introduced via a twist element ${\cal F} \in U\frv \otimes U\frv$, where $U\frv$ is the universal enveloping algebra of the Lie
algebra of vector fields $\frv:=\Gamma(TM)$ on a manifold $M$.  We use the notation ${\cal F} = \rm{f}^k\otimes \rm{f}_k$ and ${\cal F}^{-1} =  \bar{\rm{f}}^k\otimes
\bar{\rm{f}}_k$.

The invertible $\RR$-matrix $\RR\in
U\frv\otimes U\frv$ encodes the braiding and it is induced by the twist as
\begin{align}\nn
\RR=\mathcal{F}_{21}\, {\cal F}^{-1}=:\sfR^k\otimes\sfR_k \ ,
\end{align}
where $\mathcal{F}_{21}=\tau(\mathcal{F})=\mathrm{f}_k\otimes\mathrm{f}^k$ is the twist with its legs swapped. It is easy to see that the $\RR$-matrix is triangular, that is,
\begin{equation}
  \RR_{21} = \RR^{-1} = \sfR_k\otimes\sfR^k \ .\label{Inv_R}
\end{equation}

The simplest class of twists are the abelian twists, which are of the form
\begin{equation}
{\cal F} =  \exp\big(-\tfrac{\ii}{2}\,\theta^{ab}\,X_a \otimes
X_b\big) \  ,\label{Abelian_Twist}
\end{equation}
where $(\theta^{ab})$ is a constant antisymmetric matrix and $\{X_{a}\}$ is a set of commuting vector fields. For these twists, $\CF_{21}=\CF^{-1}$ and  $\RR=\CF^{-2}$. The standard Moyal twist and the angular twist of~\cite{DimitrijevicCiric:2018blz} are examples of abelian twists.

The twist (\ref{Abelian_Twist}) deforms the pointwise product $f\cdot g$ on the algebra of functions $C^\infty(M)$ to the noncommutative star-product:
\begin{align}
\begin{split}
f\star g &=\> \bar{\rm{f}}^k(f)\cdot \bar{\rm{f}}_k(g) \label{fstarg} = \sfR_k( g)\star \sfR^k( f) \\[4pt]
&=\> f\cdot g + \tfrac{\ii}{2}\,\theta^{ab}\,X_a( f) \cdot X_b (g) + O(\theta^2) \ .
\end{split}
\end{align}

Extending the action of $U\frv$ to tensor fields using the Lie derivative, the exterior algebra of differential forms $\Omega^\bullet(M)$ is deformed similarly
\begin{align}
& \omega_1\wedge_\star\omega_2 = \bar{\rm{f}}^k(\omega_1)\wedge \bar{\rm{f}}_k(\omega_2) = (-1)^{|\omega_1|\,|\omega_2|} \, \sfR_k(\omega_2)\wedge_\star \sfR^k(\omega_1) \ ,\label{StarWedge}
\end{align}
and
\begin{align}
& \diff(\omega_1\wedge_\star\omega_2) = \diff\omega_1\wedge_\star\omega_2 + (-1)^{|\omega_1|}\,
\omega_1\wedge_\star\diff \omega_2  \ .\nn
\end{align}
Here $|\omega|$ denotes the degree of a homogeneous form $\omega$. The exterior derivative $\diff$ remains undeformed. The standard integral is strictly cyclic
\begin{equation}
\int  \omega_1\wedge_\star\omega_2 = \int (-1)^{|\omega_1|\,|\omega_2|} \omega_2\wedge_\star\omega_1 .\label{CycInt}
\end{equation}

Applying the twist deformation formalism to the $L_\infty$-algebra constructed in the previous section results in the braided $L_\infty$-algebra of ECP gravity. The underlying graded vector space (\ref{eq:2.18}) remains the same, but to account for the noncommutative deformation we denote it by $V[[\hbar]]$. In particular, the brackets $\ll_n$ are deformed to twisted brackets $\ll_n^\star$: the twisted $1$-bracket is set to $\ll_1^\star = \ll_1$, while the higher brackets are defined as
\begin{align}
    \ell_n^\star (v_1,\dots,v_n)= \ell_n(v_1 \otimes_\star \cdots \otimes_\star v_n)\ ,
\end{align}
for $n\ge 2$, where $v\otimes_\star v'= \mathcal{F}^{-1}(v\otimes v')=\mathrm{f}_k(v)\otimes \mathrm{f}^k(v')$ for $v,v'\in V[[\hbar]]$. The twisted brackets $\ell_n^\star$ are braided graded symmetric, satisfying
\begin{equation}
\ell^\star_{n} (\dots, v,v',\dots) =  -(-1)^{|v|\,|v'|}\, \ell^\star_{n} (\dots, \sfR_k v',\sfR^k v,\dots) \ .\nn
\end{equation}

The non-vanishing $2$-brackets are given by
\begin{align}
\begin{split}
    \ll_2^\star \left(\begin{pmatrix}
        \xi_1\\
        \rho_1
    \end{pmatrix},
    \begin{pmatrix}
        \xi_2\\
        \rho_2
    \end{pmatrix}\right)&=\begin{pmatrix}
        [\xi_1,\xi_2]_\star\\
        \mathrm{L}_{\xi_1}^\star\rho_2 - \mathrm{L}_{\mathrm{R}_k(\xi_2)}^\star \mathrm{R}^k(
        \rho_1)+\ii[\rho_1,\rho_2]_\star
    \end{pmatrix}\in V_0\ ,\\
    \ll_2^\star \left( \begin{pmatrix}
        \xi\\
        \rho
    \end{pmatrix},
    \begin{pmatrix}
        e\\
        \omega
    \end{pmatrix}\right)&=\begin{pmatrix}
        \mathrm{L}_\xi^\star e+\ii[\rho,e]_\star\\
        \mathrm{L}_\xi^\star \omega+\ii[\rho,\omega]_\star
    \end{pmatrix}\in V_2\ ,\\
    \ll_2^\star\left(
        \begin{pmatrix}
            \xi\\
            \rho
        \end{pmatrix},
        \begin{pmatrix}
            E\\
            \Omega
        \end{pmatrix}
    \right)&=
    \begin{pmatrix}
         \mathrm{L}^\star_\xi E+\ii[\rho,E]_\star\\
         \mathrm{L}^\star_\xi \Omega+\ii[\rho,\Omega]_\star
    \end{pmatrix}\in V_2\ ,\\
  \ll_2^\star \left(\begin{pmatrix}
            e_1\\
            \omega_1
        \end{pmatrix},
        \begin{pmatrix}
            e_2\\
            \omega_2
        \end{pmatrix}\right)&=\begin{pmatrix}
            -e_1\wedge_\star \mathrm{d}\omega_2-\mathrm{R}_k(e_2)\wedge_\star \mathrm{R}^k(\mathrm{d}\omega_1)-\mathrm{d}\omega_1\wedge_\star e_2-\mathrm{R}_k(\mathrm{d}\omega_2)\wedge_\star \mathrm{R}^k(e_1)\\
            -\mathrm{d}[e_1,e_2]_\star
        \end{pmatrix}\in V_2\ ,\\
  \ll_2^\star\left(\begin{pmatrix}
        E\\
        \Omega
    \end{pmatrix} ,
    \begin{pmatrix}
        e\\
        \omega
    \end{pmatrix}
    \right)&=\begin{pmatrix}
        \mathrm{d}x^\mu \otimes \Tr ((\iota_\mu \mathrm{d}\mathrm{f}^k(e)\wedge \mathrm{f}_k(E)+\iota_\mu \mathrm{d}\mathrm{f}^k(\omega) \wedge \mathrm{f}_k(\Omega)\\
        -(\iota_\mu \mathrm{f}^k(e))\mathrm{d}\mathrm{f}_k(E)
        -(\iota_\mu \mathrm{f}^k(\omega))d\mathrm{f}_k(\Omega)))\gamma_5)\\
        \\
        -\ii E\wedge_\star e+\ii\mathrm{R}_k(e)\wedge_\star 
        \mathrm{R}^k(E)+\ii[\omega,\Omega]_\star
    \end{pmatrix}\in V_3\ ,\\
    \ll_2^\star \left(
        \begin{pmatrix}
            \xi\\
            \rho
        \end{pmatrix},
        \begin{pmatrix}
            \mathcal{X}\\
            \mathcal{P}
        \end{pmatrix}
    \right)&=
    \begin{pmatrix}
            \mathrm{L}^\star_\xi \mathcal{X} + \ii[\rho,\mathcal{X}]_\star\\
            \mathrm{L}^\star_\xi \mathcal{P}+\ii[\rho,\mathcal{P}]_\star
        \end{pmatrix}\in V_3\ .
\end{split}  
\end{align}
The braided commutators are defined as
\begin{align}
[\rho_1,\rho_2]_\star &= [\bar{\rm{f}}^k \rho_1, \bar{\rm{f}}_k \rho_2] = \rho_1 \star \rho_2 - \mathrm{R}_k \rho_2 \star \mathrm{R}^k \rho_1 \nn\\
[\xi_1,\xi_2]_\star &= [\bar{\rm{f}}^k \xi_1, \bar{\rm{f}}_k \xi_2] = \xi_1 \star \xi_2 - \mathrm{R}_k \xi_2 \star \mathrm{R}^k \xi_1 \ .\label{BrKomutatori}
\end{align}
These commutators close in the corresponding Lie algebras, $SO(1,3)$ and the Lie algebra of vector fields, respectively. Therefore, models based on NC braided symmetriesdo not require any new degrees of freedom.

The only nonzero $3$-bracket is given by
\begin{align}
\begin{split}
    &\ll_3^\star \left(
        \begin{pmatrix}
            e_1\\
            \omega_1
        \end{pmatrix},
        \begin{pmatrix}
            e_2\\
            \omega_2
        \end{pmatrix},
        \begin{pmatrix}
            e_3\\
            \omega_3
        \end{pmatrix}
        \right)\\
        &=\begin{pmatrix}\ii\big(
            e_1\wedge_\star [\omega_2,\omega_3]_\star +\mathrm{R}_k(e_2)\wedge_\star [\mathrm{R}^k(\omega_1),\omega_3]_\star+\mathrm{R}_k(e_3)\wedge_\star \mathrm{R}^k([\omega_1,\omega_2]_\star)\\
\phantom{=}+\mathrm{R}_k([\omega_2,\omega_3]_\star)\wedge_\star \mathrm{R}^k(e_1)+[\omega_1,\mathrm{R}_k(\omega_3)]_\star \wedge_\star \mathrm{R}^k(e_2)+[\mathrm{R}_k(\omega_2),\mathrm{R}^k(\omega_1)]_\star \wedge_\star e_3\\
        \phantom{=}+\frac{\Lambda}{6} \big(e_1\wedge_\star [e_2,e_3]_\star+\mathrm{R}_k(e_2)\wedge_\star [\mathrm{R}^k(e_1),e_3]_\star +\mathrm{R}_k(e_3)\wedge_\star \mathrm{R}^k([e_1,e_2]_\star)\big) \big)\\
        \\
        \ii\big(\big[\o_1, [e_2, e_3]_{\star} \big]_{\star} + \big[\mathrm{R}_k(\o_2), [\mathrm{R}^{k}(e_1), e_3]_{\star} \big]_{\star} + \big[\mathrm{R}_{k}(\o_3), \mathrm{R}^{k}([e_1, e_2]_{\star}) \big]_{\star} \big)
        \end{pmatrix}\in V_2\ .\end{split}
\end{align}
One can verify that the brackets defined above satisfy the braided homotopy relations.

\vspace{5mm}
{\bf Field equations}
\vspace{3mm}

Starting from
\begin{equation}
{\cal F}^\star_A = \sum_{n =1}^\infty \, \frac{1}{n!}\, (-1)^{\frac12\,{n\,(n-1)}}\, \ell^\star_{n}(A,\dots,A)  \nn
\end{equation}
with $A = (e,\omega)$, the braided equations of motion follow:
\begin{align}
{\cal F}_e^\star &=\frac{1}{2}\Big(e \wedge_\star \diff \o + \mathrm{R}_k(e)\wedge_\star \mathrm{R}^k(\diff \o) + \diff \o \wedge_\star e+ \mathrm{R}_k(\diff \o)\wedge_\star \mathrm{R}^k(e)\Big) \nn\\
&\phantom{=} -\frac{\ii}{6}\Big( e\wedge_\star [\omega,\omega]_\star +\mathrm{R}_k(e)\wedge_\star [\mathrm{R}^k(\omega),\omega]_\star+\mathrm{R}_k(e)\wedge_\star \mathrm{R}^k([\omega,\omega]_\star)\nn \\
& \phantom{=} +\mathrm{R}_k([\omega,\omega]_\star)\wedge_\star \mathrm{R}^k(e)+[\omega,\mathrm{R}_k(\omega)]_\star \wedge_\star \mathrm{R}^k(e)+[\o,\o]_\star \wedge_\star e \nn\\
&\phantom{=}  +\frac{\Lambda}{6} \big(e\wedge_\star [e,e]_\star+\mathrm{R}_k(e)\wedge_\star [\mathrm{R}^k(e),e]_\star +\mathrm{R}_k(e)\wedge_\star \mathrm{R}^k([e,e]_\star)\big) \Big) =0\label{BraidedEoM}\ ,
\end{align}

\begin{align}
{\cal F}_\o^\star &= \frac{1}{2}\diff [e,e]_\star \nn\\
&\phantom{=}-\frac{\ii}{6}\Big( \big[\o,[e,e]_\star \big]_\star+\big[\mathrm{R}_k(\o),[\mathrm{R}^k(e),e]_\star \big]_\star+\big[\mathrm{R}_k(\o),\mathrm{R}^k ([e,e]_\star) \big]_\star \Big)=0\label{BraidedEoM2}\ .
\end{align}

\vspace{5mm}
{\bf Cyclic pairing and action}
\vspace{3mm}

The presence of a strictly cyclic pairing
\begin{align}
& \Big\langle \begin{pmatrix}
        e\\
        \omega
    \end{pmatrix},
    \begin{pmatrix}
        E\\
        \Omega
    \end{pmatrix}
    \Big\rangle _\star = \ \int \Tr \big((e\wedge_\star E + \omega \wedge_\star \Omega)\gamma_5\big)\ , \label{BrPairing}\\
& \Big \langle
    \begin{pmatrix}
        \xi\\
        \rho
    \end{pmatrix},
    \begin{pmatrix}
        \mathcal{X}\\
        \mathcal{P}
    \end{pmatrix}
    \Big \rangle _\star
    = -\int \iota^\star_\xi \mathcal{X}-\int \Tr  (\rho \star\mathcal{P} \gamma_5)\ \nn
\end{align}
allows one to define the braided ECP action
\begin{align}
S^\star(A) = \sum_{n=1}^\infty \, \frac{1}{(n+1)!}\, (-1)^{\frac12\,{n\,(n-1)}}\, \langle A, \ell^\star_{n}(A,\dots,A)\rangle_\star \ .\nn
\end{align}
Inserting the nonzero brackets yields the explicit expression

\begin{align}
S^\star &= \frac{1}{6}\int \Tr \Big[\Big(e\wedge_\star \big(e\wedge_\star \mathrm{d}\omega+\mathrm{R}_k(e)\wedge_\star \mathrm{R}^k(\mathrm{d}\omega)+ \mathrm{d}\omega\wedge_\star e+\mathrm{R}_k(\mathrm{d}\omega)\wedge_\star \mathrm{R}^k(e)\big) \nn \\
&\phantom{=} + \o \wedge_\star \diff [e,e]_\star\Big)\gamma_5 \Big] \nn\\
& \phantom{=} -\frac{\ii}{24} \int \Tr \Big[ \Big( e \wedge_\star \big(e\wedge_\star [\omega,\omega]_\star +\mathrm{R}_k(e)\wedge_\star [\mathrm{R}^k(\omega),\omega]_\star+\mathrm{R}_k(e)\wedge_\star \mathrm{R}^k([\omega,\omega]_\star)\nn \\
&\phantom{=}+\mathrm{R}_k([\omega,\omega]_\star)\wedge_\star \mathrm{R}^k(e)+[\omega,\mathrm{R}_k(\omega)]_\star \wedge_\star \mathrm{R}^k(e)+[\o,\o]_\star \wedge_\star e \nn \\
&\phantom{=}+\frac{\Lambda}{6} (e\wedge_\star [e,e]_\star+\mathrm{R}_k(e)\wedge_\star [\mathrm{R}^k(e),e]_\star +\mathrm{R}_k(e)\wedge_\star \mathrm{R}^k([e,e]_\star))\big) \nn\\
& \phantom{=} +\o\wedge_\star \big(\big[\o,[e,e]_\star \big]_\star+\big[\mathrm{R}_k(\o),[\mathrm{R}^k(e),e]_\star \big]_\star+\big[\mathrm{R}_k(\o),\mathrm{R}^k ([e,e]_\star) \big]_\star\big)\Big) \gamma_5 \Big]\ . \label{SNonExp}
\end{align}
After evaluating the traces of the $\gamma$-matrices in (\ref{SNonExp}), we obtain:
\begin{align}
S^\star &= \int \varepsilon_{abcd} \Big[
\frac{1}{3} \left( \diff \omega^{ab} \wedge_\star e^c \wedge_\star e^d
 + e^c \wedge_\star \diff \omega^{ab} \wedge_\star e^d
 + e^c \wedge_\star e^d \wedge_\star \diff \omega^{ab}
  \right) \nn \\
 & \phantom{=}+ \frac{1}{6} \left(
 \omega^{af} \wedge_\star  {\omega_{f}}^{b}\wedge_\star e^c \wedge_\star e^d
 - \omega^{af} \wedge_\star e^c \wedge_\star {\omega_{f}}^{b} \wedge_\star e^d
 + \omega^{af} \wedge_\star e^c \wedge_\star e^d \wedge_\star  {\omega_{f}}^{b} \right. \nn \\
 &\phantom{=} \left.
 + e^c \wedge_\star \omega^{af} \wedge_\star  {\omega_{f}}^{b} \wedge_\star e^d
 - e^c \wedge_\star \omega^{af} \wedge_\star e^d \wedge_\star  {\omega_{f}}^{b}
 + e^c \wedge_\star e^d \wedge_\star \omega^{af} \wedge_\star {\omega_{f}}^{b}
  \right)\nn\\
 & \phantom{=}-\frac{\Lambda}{6} e^a \wedge_\star e^b \wedge_\star e^c \wedge_\star e^d
\Big] \nn \\
& =  \int \varepsilon_{abcd} \Big[ \frac{1}{3} \left( R^{\star ab} \wedge_\star e^c \wedge_\star e^d + e^c \wedge_\star R^{\star ab} \wedge_\star e^d + e^c \wedge_\star e^d \wedge_\star R^{\star ab}
 \right) \nn \\
& \phantom{=} + \frac{1}{6} \left(
 - \omega^{af} \wedge_\star  {\omega_{f}}^{b}\wedge_\star e^c \wedge_\star e^d
 - \omega^{af} \wedge_\star e^c \wedge_\star {\omega_{f}}^{b} \wedge_\star e^d
 + \omega^{af} \wedge_\star e^c \wedge_\star e^d \wedge_\star  {\omega_{f}}^{b} \right. \nn \\
 & \phantom{=}\left.
 - e^c \wedge_\star \omega^{af} \wedge_\star  {\omega_{f}}^{b} \wedge_\star e^d
 - e^c \wedge_\star \omega^{af} \wedge_\star e^d \wedge_\star  {\omega_{f}}^{b}
 - e^c \wedge_\star e^d \wedge_\star \omega^{af} \wedge_\star {\omega_{f}}^{b}
  \right)\nn\\
 & \phantom{=}-\frac{\Lambda}{6} e^a \wedge_\star e^b \wedge_\star e^c \wedge_\star e^d
\Big], \label{SNonExpTrace}
\end{align}
where
\begin{equation*}
R^{\star} = \diff \omega - \frac{\ii}{2} [\omega, \omega]_{\star}
\end{equation*}
is the NC curvature.

Comparing (\ref{SNonExpTrace}) to (\ref{eq:2.6}), one sees that the braided action can be obtained from the commutative one by substituting commutative wedge products with noncommutative ones, while doing so for all commutatively equivalent orderings of the fields with equal weights. A straightforward calculation shows that the action (\ref{SNonExp}) is indeed invariant under the braided diffeomorphism and the braided $SO(1,3)$ infinitesimal transformations
\begin{align}
\d^\star_{\Big( {\substack{\x \\ \r}}\Big)}\begin{pmatrix}
        e\\
        \o
    \end{pmatrix}
    =\ell_1^\star \bpm \xi\\ \rho \epm+\ell_2^\star \bpm\bpm \xi \\ \rho \epm, \bpm e\\ \o \epm \epm .\label{BrSymm}
\end{align}

Finally, using the strict cyclicity of the integral (\ref{CycInt}) the action (\ref{SNonExpTrace}) can be simplified to
\begin{align}
S^\star &=   \int \varepsilon_{abcd} \Big[ R^{\star ab} \wedge_\star e^c \wedge_\star e^d -\frac{\Lambda}{6} e^a \wedge_\star e^b \wedge_\star e^c \wedge_\star e^d \nn\\
& \phantom{=}- \frac{1}{3} \big( \omega^{af} \wedge_\star  {\omega_{f}}^{b}\wedge_\star e^c \wedge_\star e^d
 + \omega^{af} \wedge_\star e^c \wedge_\star {\omega_{f}}^{b} \wedge_\star e^d \big) \Big] \ . \label{SNonExpTraceCycl}
\end{align}

In the next section we will discuss some phenomenological implications of equations (\ref{BraidedEoM})-(\ref{BraidedEoM2}) and the braided action (\ref{SNonExp}). Although more general deformations can be considered, for convenience we choose to work with the Moyal twist $\mathcal{F}$
\begin{equation}
{\cal F} =  e^{-\frac{\ii\hbar}{2}\theta^{\mu\nu}\partial_\mu \otimes
\partial_\nu} .\label{MoyalTwist}
\end{equation}
The small parameter $\hbar$ and the constant antisymmetric matrix $\theta^{\mu\nu}$ are often written together as $\theta^{\mu\nu}$ and we refer to the components  $\theta^{\mu\nu}$
  as the small noncommutative deformation parameters.

\counterwithin*{equation}{section}
\section{Second order expansion}

As mentioned at the beginning of our paper, our aim is to better understand the physical implications of the braided ECP gravity and to compare this model with other NC gravity models found in the literature. To discuss phenomenological implications of NC gravity, different directions can be pursued. In this paper, we focus on:

\begin{itemize}

\item[$\bullet$] Searching for  specific solutions of the NC gravity equations of motion and analyzing their properties.

\item[$\bullet$] Expanding the NC gravity action and identifying additional terms induced by noncommutativity in the three-graviton vertex.

\end{itemize}

\subsection{Braided field equations}

Finding exact gravitational solutions is challenging even  in  commutative General Relativity. Therefore, we first expand equations (\ref{BraidedEoM}) and (\ref{BraidedEoM2}) up to second order in the NC parameter $\theta^{\alpha\beta}$ and search for solutions perturbatively, expanding around a known solution of the commutative theory. Note that the first-order NC correction vanishes, as expected.

The second-order expanded equation of motion for the vierbein is given by
\begin{align}\label{eq:4.1}
\begin{split}
\mathcal{F}_e^\star &= e\wedge R+R\wedge e+\frac{\ii \Lambda }{3}e\wedge e \wedge e\\
&\phantom{=}+\frac{\ii}{24}\theta^{\mu\nu}\theta^{\rho\sigma}\Big(3\ii(\partial_\mu\partial_\rho e\wedge \partial_\nu \partial_\sigma \mathrm{d}\omega+\partial_\mu \partial_\rho \mathrm{d}\omega \wedge \partial_\nu \partial_\sigma e)\\
&\phantom{=}+3(e\wedge \partial_\mu \partial_\rho \omega \wedge \partial_\nu \partial_\sigma \omega+\partial_\mu \partial_\rho \omega \wedge \partial_\nu \partial_\sigma \omega \wedge e)\\
&\phantom{=}+2(\partial_\mu e \wedge [\partial_\nu \partial_\sigma \omega,\partial_\rho \omega]-[\partial_\rho \omega,\partial_\mu\partial_\sigma \omega]\wedge \partial_\nu e)\\
&\phantom{=}+3(\partial_\mu \partial_\rho e \wedge [\omega,\partial_\nu \partial_\sigma \omega]+[\omega,\partial_\mu\partial_\rho \omega]\wedge \partial_\nu \partial_\sigma e)\\
&\phantom{=}+ \partial_\mu \partial_\rho e \wedge [\partial_\nu \omega,\partial_\sigma \omega]+[\partial_\mu,\partial_\rho \omega]\wedge \partial_\nu \partial_\sigma e\Big)\\
&\phantom{=}-\frac{\ii \Lambda}{24}\theta^{\m\n}\theta^{\r\s}\bigg(-\frac{1}{2}e\wedge\partial_\mu \partial_\rho e\wedge \partial_\nu \partial_\sigma e-\frac{1}{4}\partial_\mu \partial_\rho e \wedge \partial_\nu \partial_\sigma [e,e]\\
&\phantom{=}+\frac{1}{3} \partial_\mu e\wedge \partial_\rho \partial_\nu e \wedge \partial_\sigma e -\frac{1}{3}\partial_\mu e \wedge \partial_\rho e \wedge \partial_\sigma \partial_\nu e\\
&\phantom{=}+\frac{1}{3} \partial_\mu \partial_\rho e \wedge \partial_\sigma [\partial_\nu e,e]-\frac{1}{3}\partial_\mu \partial_\rho e \wedge [\partial_\nu \partial_\sigma e,e]\bigg)\ .
\end{split} 
\end{align}
The second-order expanded equation of motion for the spin connection is
\begin{align} \label{eq:4.2}
    \begin{split}
        \mathcal{F}_\omega ^\star &=T\wedge e - e\wedge T\\
        &\phantom{=}+\frac{\ii}{24}\theta^{\mu\nu}\theta^{\rho\sigma}\Big(-3\ii[\pa_\m \pa_\r e,\pa_\n \pa_\s \mathrm{d}e]
        +3 \big[\pa_\m \pa_\r \o,[\pa_\n \pa_\s e, e]\big]\\
    &\phantom{=}+ \big[\partial_\mu \pa_\r \o, [\pa_\n e,\pa_\s e] \big]
    +\frac{3}{2} \big[\o, [\pa_\m \pa_\r e,\pa_\n \pa_\s e] \big] + 2 \big[ \pa_\m \o, [\pa_\r e, \pa_\n \pa_\s e]\big]\Big)\ .
    \end{split}
\end{align}

Expanding the Lie algebra-valued vierbein and spin-connection in the coordinate basis, field equations \eqref{eq:4.1} and \eqref{eq:4.2} can be recast in the more familiar form:
\begin{align}
    \label{eq:4.3}
    {R_d}^\sigma -\frac{1}{2}{e_d}^\sigma R&={N_d}^\sigma\ ,\\
    \label{eq:4.4}
    {T^a}_{ac} {e_d}^\delta - {T^a}_{ad} {e_c}^\delta +{T^\d}_{cd} &={S_{cd}}^\delta\ .
\end{align}
The expressions ${S_{cd}}^\delta$ and ${N_d}^\s$ are the second-order braided noncommutative corrections to the undeformed (commutative) field equations with the vierbein determinant denoted as $|e|$, and they are given by
\begin{align}
\begin{split}  \label{eq:4.5}
    {N_d}^\s &= \frac{1}{8}\theta^{\a\b}\theta^{\g\d} \Bigg[ \frac{1}{2}(\pa_\a \pa_\g {e^a}_\m) (\pa_\b \pa_\d \pa_\n  {\o^{bc}}_\r) + \frac{1}{2} {e^a}_\m (\pa_\a \pa_\g {\o^{bf}}_\n) (\pa_\b \pa_\d {{\o_f}^c}_\r) \\
&\phantom{=}+ \frac{2}{3} (\pa_\a {e^a}_\m) (\pa_\b \pa_\d {\o^{bf}}_\n) (\pa_\g {{\o_f}^c}_\r) + (\pa_\a \pa_\g {e^a}_\m) (\pa_\b \pa_\d  {\o^{bf}}_\n) {{\o_f}^c}_\r \\
&\phantom{=}+ \frac{1}{3} (\pa_\a \pa_\g {e^a}_\m) (\pa_\b  {\o^{bf}}_\n) (\pa_\d {{\o_f}^c}_\r)\Bigg]
\frac{1}{|e|}
\varepsilon^{\mu\nu\rho\sigma}
\varepsilon_{abcd} \ ,
\end{split}
\end{align}
and
\begin{align} 
\begin{split} \label{eq:4.6}
{S_{cd}}^\s &= \frac{1}{8} \theta^{\a\b}\theta^{\g\d} \Bigg[(\pa_\a \pa_\g {e^a}_\m)(\pa_\b \pa_\d \pa_\r {e^b}_\n) +
(\pa_\a \pa_\g {\o^{ac}}_\m)(\pa_\b \pa_\d e_{c\n}) {e^b}_\r
+ (\pa_\a \pa_\g {\o^{ac}}_\m) e_{c\n} (\pa_\b \pa_\d {e^b}_\r)  \\
&\phantom{=} + {\o^{ac}}_\m (\pa_\a \pa_\g e_{c\n}) (\pa_\b \pa_\d {e^b}_\r)
+ \frac{2}{3} (\pa_\a \pa_\g {\o^{ac}}_\m) (\pa_\b
e_{c\n}) (\pa_\d {e^b}_\r)  \\
&\phantom{=} +  \frac{2}{3} (\pa_\b {\o^{ac}}_\m) (\pa_\a \pa_\g
e_{c\n}) (\pa_\d {e^b}_\r) + \frac{2}{3} (\pa_\a {\o^{ac}}_\m) (\pa_\g
 e_{c\n}) (\pa_\b \pa_\d {e^b}_\r)
\Bigg]\frac{1}{|e|}\varepsilon^{\m\n\r\s} \varepsilon_{abcd}\ .
\end{split}
\end{align}
From \eqref{eq:4.3} and \eqref{eq:4.4}, it is clear that the presence of noncommutativity induces nonzero curvature and torsion. Furthermore, torsion can be expressed in terms of the braided noncommutative correction ${S_{cd}}^\d$ as follows. First, we multiply both sides of \eqref{eq:4.4} by ${e^b}_\d$; by contracting local indices, we obtain the partial trace of the torsion:
\begin{align}
    {T^a}_{ad}=-\frac{1}{2}{S_{ad}}^a\ ,
\end{align}
where ${S_{ab}}^c={e^c}_\mu {S_{ab}}^\mu$. Then, the braided torsion equation can be rewritten as
\begin{align}
    {T^b}_{cd}={S_{cd}}^b +\frac{1}{2}\left(\d^b_d {S_{ac}}^a - \d^b_c {S_{ad}}^a\right)\ , \label{eq:4.7}
\end{align}
so that in the undeformed theory, we recover the zero torsion condition, as expected.

\vspace{5mm}
{\bf Braided NC Minkowski spacetime}
\vspace{3mm}

In the commutative theory, Minkowski spacetime is a vacuum solution of the Einstein equations without a cosmological constant. Thus, we set $\Lambda = 0$  from now on. The braided NC Minkowski spacetime can be expressed perturbatively up to second order in the NC parameter $\theta^{\alpha\beta}$ as
\begin{align}
    e=e^{(0)}+e^{(1)}+e^{(2)}\ ,
\end{align}
where $e^{(0)}=\d^a_\mu \g_a \mathrm{d}x^\m$ is the vierbein $1$-form corresponding to commutative Minkowski spacetime, while $e^{(1)}$ and $e^{(2)}$ are the first- and second-order braided NC corrections, respectively, to be determined by solving equation (\ref{eq:4.3}). At zeroth order, equation \eqref{eq:4.3} reduces to the Einstein equation, for which $e^{(0)}$ is an exact solution. The right-hand side vanishes, since the partial derivatives act only on the (constant) commutative part of the vierbein $e^{(0)}$. We thus find that there are no second-order braided NC corrections to the field equations in the case of the braided NC Minkowski spacetime, i.e., commutative Minkowski spacetime remains a solution of the braided field equations. It follows from these equations that the curvature vanishes up to second order in the NC parameter. As seen from \eqref{eq:4.7}, the second-order correction to the torsion also vanishes, so that the second-order spin connection retains only its unique torsion-free part.

The Moyal twist (\ref{MoyalTwist}) acts trivially on the geometry of Minkowski spacetime; in particular
\begin{equation}
{\rm d}x^\mu \wedge_\star {\rm d}x^\nu = {\rm d}x^\mu\wedge {\rm d}x^\nu, \quad f\star {\rm d}x^\mu ={\rm d}x^\mu\star f = f\cdot {\rm d}x^\mu,\label{Pomocne}
\end{equation}
for an arbitrary function $f$. Using this property, a stronger conclusion follows from the braided equations of motion (\ref{BraidedEoM}) and (\ref{BraidedEoM2}): Commutative Minkowski spacetime is an exact solution of braided NC gravity.

This result is in agreement with the model of twisted NC gravity \cite{Aschieri:2005yw}. In the NC gravity model based on the $SO(1,3)_\star$ gauge symmetry \cite{Aschieri:2011ng}  commutative Minkowski spacetime is also an exact solution. However, in \cite{DimitrijevicCiric:2016qio} a different result was obtained. There, a model based on the $SO(2,3)_\star$ gauge NC gravity was constructed, and it was shown that NC corrections induce curvature and torsion in Minkowski spacetime. These NC corrections appear at second order in the NC parameter $\theta^{\alpha\beta}$.

\vspace{5mm}
{\bf Braided de Sitter space}
\vspace{3mm}

The flat slicing of de Sitter (dS) space, which covers half of the full de Sitter space, is given by
\begin{align}
    \diff s^2=\diff t^2-e^{\frac{2t}{\a}}(\diff x^i)^2\ .
\end{align}
A choice of vierbein fields can be realized as
\begin{align}
    {e^a}_\m=\d_\m^0 \d^a_0 + \d_\m^i \d^a_i e^{\frac{t}{\a}}\ .
\end{align}
Similarly to Minkowski spacetime, the flat dS geometry is invariant under the action of the Moyal twist, and the relations (\ref{Pomocne}) hold. It then follows that flat dS is an exact solution of the braided equations (\ref{BraidedEoM}) and (\ref{BraidedEoM2}).

\subsection{Three-graviton vertex}

Let us now consider the second-order NC correction to the three-graviton vertex in braided NC gravity. Following \cite{Alvarez-Gaume:2006qrw}, we start with the weak field expansion around flat spacetime. Then the metric $g_{\m\n}$ is given by
\begin{align}
    g_{\m\n}=\eta_{\m\n}+2\k h_{\m\n}\ , \label{MetricPerturbation}
\end{align}
where $\k$ is the perturbation parameter. In terms of the vierbein, this expansion is
\begin{align}
{e^a}_\m={\d^a}_\m + \k {\t^a}_\m ,\quad {\t^a}_\m = \eta^{ab} h_{\mu\nu} {\d_b}^{\n}  \ . \label{VierbeinExp}
\end{align}
The vierbein of the flat metric is ${\d^a}_\m$, with
\begin{align}
\begin{split}
    \eta_{\mu\nu}={\d^a}_\m {\d^b}_\n \eta_{ab}\ , \quad 
   \d^\m_\n= {\d_a}^\m {\d^a}_\n \ , \quad 
   \d^b_a={\d_a}^\m {\d^b}_\m\ .
\end{split} \nn
\end{align}
The first-order perturbation of the flat vierbein ${\d^a}_\m$ is denoted by ${\t^a}_\m$. Since the quantity of interest is the three-graviton vertex, the inverse vierbein should be expanded to second order in the perturbation parameter $\k$:
\begin{align}
    {e_a}^\m={\d_a}^\m +\k {\t_a}^{\m(1)}+\k^2 {\t_a}^{\m (2)}\ , \nn
    \end{align}
so that
\begin{align}
\begin{split}
{\t_a}^{\m (1)}&=- {\d_b}^\m {\d_a}^\n {\t^b}_\n\ ,\\
{\t_a}^{\m (2)}&={\d_c}^\m {\d_b}^\n  {\d_a}^\r {\t^c}_\n {\t^b}_\r\ . \label{InvTetradaRazvoj}
\end{split}
\end{align}

The linearized theory is invariant under the infinitesimal gauge transformations
\begin{equation}
\delta_\xi \k h_{\mu\nu} = \k\partial_\mu\xi_\nu + \k\partial_\nu\xi_\mu \nn
\end{equation}
with an arbitrary vector field $\xi$. This symmetry must be gauge-fixed in order to obtain a valid propagator for the gravitational field. We choose the  transverse-traceless gauge:
\begin{equation}
\partial_\mu h^{\mu\nu} =0 , \quad h^\mu_{\ \ \mu} =0 \  \label{GaugeFixing}
\end{equation}
and perform our calculations using this gauge choice. The action (\ref{SNonExpTrace}) can be expanded in powers of the NC parameter $\theta^{\alpha\beta}$, with the first nontrivial NC correction quadratic in $\theta^{\alpha\beta}$ given by
\begin{align} \label{eq:4.10}
S^\star&= S^{(0)} + S^{(2)} \nn\\  
    &= \int \varepsilon_{abcd} R^{ab}\wedge e^c \wedge e^d +\frac{1}{8} \varepsilon_{abcd}\theta^{\a\b}\theta^{\g\d}\Bigg( -\frac{2}{3}\pa_\a \pa_\g \mathrm{d}\o^{ab}\wedge \pa_\b \pa_\d e^c\wedge e^d \nn \\
    &\phantom{=}-\frac{1}{3}\mathrm{d}\o^{ab}\wedge \pa_\a \pa_\g e^c \wedge \partial_\b \partial_\d e^d
    -\frac{1}{2}\partial_\a \partial_\g \o^{ae}\wedge \partial_\b \partial_\d {\o_e}^b \wedge e^c\wedge e^d \nn \\
    &\phantom{=}-\frac{2}{3}\pa_\a \pa_\g \o^{ae}\wedge \pa_\d {\o_e}^b \wedge \pa_\b e^c \wedge e^d-2\pa_\a \pa_\g \o^{ae}\wedge {\o_e}^b \wedge \pa_\b \pa_\d e^c\wedge e^d \nn \\
    &\phantom{=}-\frac{1}{3}\pa_\a \pa_\g \o^{ae}\wedge {\o_e}^b \wedge \pa_\b e^c \wedge \pa_\d e^d
    -\frac{1}{3}\pa_\a \o^{ae}\wedge\pa_\g {\o_e}^b \wedge \pa_\b \pa_\d e^c\wedge e^d \nn \\
    &\phantom{=}-\frac{2}{3} \pa_\a \o^{ae}\wedge {\o_e}^b \wedge \pa_\b \pa_\d e^c \wedge \pa_\g e^d 
    -\frac{1}{2}\o^{ae}\wedge {\o_e}^b\wedge \pa_\a \pa_\g e^c \wedge \pa_\b \pa_\d e^d \Bigg)\ .
\end{align}

The expanded action (\ref{eq:4.10}) is a function of independent fields ${e^a}_\m$ and ${\o^{ab}}_\m$. The lowest-order NC correction to the
three-graviton vertex is expected to be third order in the perturbation parameter $\k$ and quadratic in the NC parameter $\theta^{\a\b}$, and is most commonly given in terms of the metric perturbation $h_{\mu\nu}$. Therefore, we express the spin connection in terms of the vierbein (and torsion) and then rewrite the relevant terms in the expansion in terms of $h_{\mu\nu}$.

The low-energy expansion of the vierbein is given by (\ref{VierbeinExp}). Let us discuss the  expansion order of the spin connection ${\o^{ab}}_\m$. In the undeformed gravity, the sipn connection is a function of the vierbien ${e^a}_\m$ and contorsion $K_{\m[\nu\rho]}$ \cite{Blagojevic:2002du, Freedman:2012zz} \footnote{The (anti)symmetrization of indices is done with weight $1/2$
\begin{equation}
A_{[\mu\nu]} = \frac{1}{2}(A_{\mu\nu} - A_{\nu\mu}), \quad\quad A_{(\mu\nu)} = \frac{1}{2}(A_{\mu\nu} + A_{\nu\mu}) .\nn 
\end{equation}
}
\begin{align}
\o_{\m[\nu\rho]} &=  {\o^{ab}}_\m (e) {e_a}^\nu {e_a}^\rho + K_{\m[\nu\rho]},\nn\\
{\o^{ab}}_\m (e) &= 2 e^{\nu[a}\partial_{[\mu} e_{\nu]}^{b]} - e^{\nu[a}e^{b]\sigma}e_{\mu c}\partial_\nu{e_\sigma}^c,\label{TorsionFreeOmega}\\
K_{\m[\nu\rho]} &= -\frac{1}{2}\Big( T_{[\m\nu]\rho} - T_{[\nu\rho]\mu} + T_{[\rho\m]\nu}  \Big) \>{\mbox{ and }}\> T_{[\mu\nu]\rho} = {T_{\mu\nu}}^a {e_a}^\rho .\label{Contorsion}  
\end{align}
In our braided NC gravity model, the torsion does not vanish (\ref{eq:4.7}); therefore, the contorsion contribution should be included in the spin connection. However, as the braided NC correction to the torsion is already of second order in the deformation parameter $\theta^{\a\b}$ (cf. \eqref{eq:4.7}), and the action functional is expanded to the same order, only the torsion-free part of the spin connection (\ref{TorsionFreeOmega}) contributes to the expansion defiend above. In that case, the spin connection can be expanded in powers of the perturbation parameter $\k$ as
\begin{align} \label{eq:4.23}
{\o^{ab}}_\m=\o^{ab(0)}_{\enspace \m}+\k \o^{ab(1)}_{\enspace \m}+\k^2 \o^{ab(2)}_{\enspace \m}\ .
\end{align}
Since we are expanding around flat spacetime, the zeroth-order term vanishes:
\begin{equation}
\o^{ab(0)}_{\enspace \m} =0 \ .\nn 
\end{equation}
As we will see, the only nonvanishing contribution to the braided NC correction of the three-graviton vertex arises from the term linear in $\k$:
\begin{align}
\o^{ab(1)}_{\enspace \m}=\delta^a_\r \delta^b_\s (\pa^\s {h_\m}^\r-\pa^\r {h_\m}^\s)\ , \label{Omega1}
\end{align}
where ${h_\m}^\nu=\d^\r_a \d^\nu_b \eta^{ab}h_{\mu\rho}$ and the explicit expression (\ref{Omega1}) follows from (\ref{TorsionFreeOmega}).

With the the field content defined, we now extract the part of the action \eqref{eq:4.10} that contributes to the three-graviton vertex and is of leading (second) order in the deformation parameter $\theta^{\mu\nu}$:
\begin{align*}
\begin{split}
    S^\star_{3,\theta^2}&=\frac{\k^3}{8}\q^{\a\b}\q^{\g\d}\int \mathrm{d}^4 x \varepsilon^{\m\n\r\s} \varepsilon_{abcd} \bigg[\\
 &-\frac{2}{3}\bigg(\left(\pa_\a \pa_\g \pa_\m {\o^{ab}}^{(1)}_\n\right)\left(\pa_\b \pa_\d {\t^c}_\r\right) {\t^d}_\s
    +\left(\pa_\a \pa_\g \pa_\m {\o^{ab}}^{(2)}_\n\right)\left(\pa_\b \pa_\d {\t^c}_\r\bigg) {\d^d}_\s\right)\\
    &\phantom{=}-\frac{1}{3}\left(\pa_\m {\o^{ab}}^{(1)}_\n \right)\left(\pa_\a \pa_\g {\t^c}_\r  \right)\left(\pa_\b \pa_\d {\t^d}_\s \right)
    - \left(\pa_\a \pa_\g {\o^{ae}}^{(1)}_\m \right)\left(\pa_\b \pa_\d {{\o_e}^b}^{(1)}_\n \right){\t^c}_\r {\d^d}_\s\\
    &\phantom{=}-\left(\pa_\a \pa_\g {\o^{ae}}^{(2)}_\m \right)\left(\pa_\b \pa_\d {{\o_e}^b}^{(1)}_\n  \right){\d^c}_\r {\d^d}_\s
    -\frac{2}{3}\left(\pa_\a \pa_\g {\o^{ae}}^{(1)}_\m \right)\left( \pa_\d {{\o_e}^b}^{(1)}_\n  \right)\left(\pa_\b{\t^c}_\r \right){\d^d}_\s\\
    &\phantom{=}-2\left(\pa_\a \pa_\g {\o^{ae}}^{(1)}_\m \right)\left({\pa_\b \pa_\d \t^c}_\r\right) {{\o_e}^b}^{(1)}_\n {\d^d}_\s
    -\frac{1}{3}\left(\pa_\a  {\o^{ae}}^{(1)}_\m \right)\left( \pa_\g{{\o_e}^b}^{(1)}_\n  \right)\left({\pa_\b \pa_\d \t^c}_\r\right) {\d^d}_\s\bigg] \ .
\end{split}
\end{align*}
The terms involving the second-order correction to the spin connection ${\o^{ab}}^{(2)}_\mu$ vanish due to the antisymmetry of the deformation parameter $\theta^{\mu\nu}$ and the boundary behaviour of the fields. Consequently, the relevant contribution reduces to
\begin{align}
    S^\star_{3,\theta^2}&=\frac{\k^3}{2}\theta^{\a\b}\theta^{\g\d} \int \mathrm{d}^4 x \Big[2(\pa_\a \pa_\g h^{\mu\rho})(\pa_\b \pa_\d h^{\n\s})(\pa_\m \pa_\n h_{\r\s})-(\pa_\a \pa_\g h^{\m\n})(\pa_\b \pa_\d h^{\r\s})(\pa_\m \pa_\n h_{\r\s}))\Big]\ . \label{3VertexResult}
\end{align}

Surprisingly, the result (\ref{3VertexResult}) is identical to the three-graviton vertex obtained in \cite{Alvarez-Gaume:2006qrw}. Therefore, the twisted gravity from \cite{Aschieri:2005yw} and the braided gravity from \cite{DimitrijevicCiric:2021jea} agree at the level of the three-graviton vertex. However, the result differs from that obtained in the NC gravity model induced by low-energy string theory \cite{Alvarez-Gaume:2006qrw}.

\section{Discussion}

In this paper, we discussed some phenomenological implications of braided NC gravity in four dimenstions. We first rewrote the corresponding braided $L_\infty$-algebra in a more standard $\gamma$-matrix basis. Choosing the Moyal twist to introduce the NC deformation, we showed that Minkowski spacetime and flat de Sitter spacetime are exact solutions of the vacuum braided equations of motion. As the next step, the three-graviton vertex was calculated and compared with the results obtained in \cite{Alvarez-Gaume:2006qrw}. The result obtained is identical to the result from the twisted gravity model \cite{Aschieri:2005yw}.

One should bear in mind that the twisted gravity and braided NC gravity models are not the same. In the twisted gravity model \cite{Aschieri:2005yw} a NC deformation of the Einstein-Hilbert action is constructed, while the braided NC gravity \cite{DimitrijevicCiric:2021jea} is a defomation of the more general Einstein-Cartan-Palatini gravity.
Additionally, the twisted gravity is based on twisted diffeomorphism invariance, and the construction of its action and equations of motion is not unique. On the other hand braided NC gravity is based on the corresponding braided $L_\infty$-algebra that uniquely defines both the symmetries of the model and the dynamics, including the action and the equations of motion. The braided NC gravity action and equations of motion differ from the corresponding twisted gravity action and equations of motion.  It would be intersting to compare different solutions in these models, in particular following \cite{Ohl:2008tw, Ohl:2009pv}, and understand their differences and/or similarities.

Another line of future research is to investigate whether the $SO(1,3)_\star$ and $SO(2,3)_\star$ NC gauge models of gravity can reproduce the string theory results for the NC gravity correction to the three-graviton vertex obtained in \cite{Alvarez-Gaume:2006qrw}. Additionally, the new double copy result for the three-graviton interaction \cite{Jonke:2025pkj} can be compared with these existing results. Understanding relations between results obtained from different NC gravity models could help connect various approaches and take first steps toward a more unified NC gravity picture.

\subsection*{Acknowledgements}

This work is supported by Project 451-03-136/2025-03/200162 of the Serbian Ministry of Science,
Technological Development and Innovation. The authors also acknowledge the support from the COST (European Cooperation in Science and Technology) Actions CaLISTA CA21109.


\appendix

\renewcommand{\theequation}{\Alph{section}.\arabic{equation}}
\setcounter{equation}{0}

\section{Gamma matrices in four dimensions}

\label{appendix:a}

Gamma matrices generate a representation of the Clifford algebra and satisfy the following anticommutation relations:
\begin{align}
\{\gamma_a, \gamma_b\} = 2 \eta_{ab} , 
\end{align}
where $\eta_{ab}$ is the metric of the Minkowski spacetime.

Any matrix $A\in GL(4,\mathbb{C})$ can be written in terms of the following set of matrices constructed from the gamma matrices \cite{Radovanovic:2008zz}: $\{I_4, \g^a, \g^5, \g^a \g^5, \Sigma^{ab} \}$, where $I_4$ is the identity $4\times 4$ matrix, the matrix $\g^5$ is defined as
\begin{align}
    \g^5=\ii \g^0 \g^1 \g^2 \g^3\ ,
\end{align}
and it satisfies $\{\g^5,\g^a \}=0$. The Lorentz generators $\S_{ab}$ are defined as
\begin{align}
    \Sigma_{ab}=\frac{\ii}{4}[\gamma_a, \gamma_b]\ ,
\end{align}
so that the corresponding commutation relations are
\begin{align}
     [\Sigma_{ab}, \Sigma_{cd}] =  \ii \left(\eta_{bc} \Sigma_{ad} + \eta_{ad} \Sigma_{bc} - \eta_{ac} \Sigma_{bd} - \eta_{bd} \Sigma_{ac}\right)\ .
\end{align}
In the following, a list of useful identities regarding gamma matrices is given:
\begin{align}
\begin{split}
& \gamma_{a} \gamma_{b} = \eta_{ab} - 2\ii \Sigma_{ab} \ , \\
& [\gamma_a, \Sigma_{bc}] = -  \ii (\eta_{ac} \gamma_{b} - \eta_{ab} \gamma_{c})\ , \\
& \{\gamma_{a}, \Sigma_{bc} \} =  \varepsilon_{abcd} \gamma_{5} \gamma^{d} \ , \\
& \{\Sigma_{ab}, \Sigma_{cd} \} = \frac{1}{2}\left( \ii \varepsilon_{abcd} \gamma_{5} +  \eta_{ac} \eta_{bd} -  \eta_{ad} \eta_{bc} \right)\ , \\
& \gamma_{a} \gamma_{b} \gamma_{c} = \eta_{bc} \gamma_{a} - \eta_{ac} \gamma_{b} + \eta_{ab} \gamma_{c} - \ii \varepsilon_{abcd} \gamma_{5} \gamma^{d} \ , \\
& \gamma_{[a} \gamma_{b} \gamma_{c]} = - \ii \varepsilon_{abcd} \gamma_{5} \gamma^{d} \ , \\
& \gamma_{[a} \gamma_{b} \gamma_{c} \gamma_{d]} = -  \ii \varepsilon_{abcd} \gamma_5 \ , \\ 
&\Tr(\gamma_{a} \gamma_{b}) = 4 \eta_{ab} \ , \\
&\Tr\left(\gamma_5 \Sigma_{ab} \Sigma_{cd}\right) =  \ii \varepsilon_{abcd}\ ,\\
&\Tr \left(\g_5\g_a \g_b \g_c \g_d \right)=-4\ii \varepsilon_{abcd}\ ,\\
&\g_a^\dagger=\g_0 \g_a \g_0\ ,\\
&\Sigma_{ab}^\dagger=\g_0 \Sigma_{ab} \g_0\ .
\end{split}
\end{align}

\section{Homotopy relations}

\label{appendix:b}

The $n$-brackets $\ll_n$ of an $L_{\infty}$ algebra are required to satisfy homotopy relations $\mathcal{J}_{n}(v_1, v_2, \dots, v_n)= 0$ for every $n\geq 1$ :

\bea
&&\mathcal{J}_{n}(v_1,\dots, v_n) = \sum_{i=1}^{n}(-1)^{i(n-i)} \sum_{\sigma \in \mathrm{Sh}_{i, n-i}}\chi(\sigma; v_1, \dots, v_n) \nn\\
&&\hskip 4cm \times\ll_{n+1-i}(\ll_{i}(v_{\sigma(1)}, \dots,  v_{\sigma(i)}),  v_{\sigma(i+1)}, \dots, v_{\sigma(n)}) ,
\eea
where the second sum runs over $(i, n-1)$-shuffled permutations $\sigma\in \mathrm{S}_n$ which satisfy
\be
\sigma(1)< \dots <\sigma(i) , \hskip 1cm \sigma(i+1)< \dots < \sigma(n) . \nn
\ee

The Koszul sign $\chi(v_1, \dots, v_n) = \pm 1$ is determined from the grading by
\be
v_{\sigma(1)} \wedge \dots \wedge v_{\sigma(n)} = \chi(\sigma, v_1,\dots,v_n) v_1\wedge \dots \wedge v_n  . \nn
\ee

The non-trivial homotopy relations of the 4D ECP gravity $L_\infty$-algebra are $\mathcal{J}_{n}$ for $n=1,\dots, 4$, because higher order homotopy relations include in every summand at least one $n$-bracket with $n\geq4$.
The homotopy relation $\mathcal{J}_{n}$ has degree $|\mathcal{J}_{n}| = |\ll_{n+1-i}| + |\ll_{i}| = 3-n$, and the underlying vector space of this $L_\infty$-algebra has degrees ranging form $0$ to $3$. By counting the total degree of the expression $\mathcal{J}_{n}(v_1,\dots, v_n)$, one can further restrict the non-trivial homotopy relations that need to be verified.

For $n=1$, one only needs to check $\mathcal{J}_{1}(v)=\ll_{1}(\ll_{1}(v)) = 0$, for $v\in V_0$ and $v\in V_1$. This homotopy relation corresponds to the differential condition for $\ll_1$.

The homotopy relation for $n=2$ expresses the Leibniz rule for the action of the differential $\ll_1$ on the $2$-bracket $\ll_2$, while the homotopy relation $\mathcal{J}_3$ is the generalized Jacoby identity for the $2$-bracket $\ll_2$.

By considering all possible combinations of arguments of the homotopy relations, one can verify their validity through explicit substitution of the expressions for the $n$-brackets.

As an example, we can prove the homotopy relation $\mathcal{J}_3$ when acting on one pair of gauge parameters of degree $0$ and two pairs of dynamical fields of degree $1$:
{\small
\begin{eqnarray}
&& \mathcal{J}_{3} \left( \begin{pmatrix} \xi\\ \rho \end{pmatrix} , \begin{pmatrix} e_1\\ \omega_1 \end{pmatrix}, \begin{pmatrix} e_2\\ \omega_2 \end{pmatrix}  \right)  \nn \\
&&  = \ll_3 \left( \ll_1 \left( \begin{pmatrix} \xi\\ \rho \end{pmatrix} \right),
\begin{pmatrix} e_1\\ \omega_1 \end{pmatrix}, \begin{pmatrix} e_2\\ \omega_2 \end{pmatrix} \right) +
\ll_2 \left( \ll_2
   \left(\begin{pmatrix}
   \xi \\ \rho \end{pmatrix}, \begin{pmatrix}
   e_1 \\ \omega_1 \end{pmatrix}\right), \begin{pmatrix}
   e_2 \\ \omega_2 \end{pmatrix} \right)   \nonumber \\
   && + \ll_2 \left( \ll_2
    \left(\begin{pmatrix}
   \xi \\ \rho \end{pmatrix}, \begin{pmatrix}
   e_2 \\ \omega_2 \end{pmatrix}\right), \begin{pmatrix}
   e_1 \\ \omega_1 \end{pmatrix} \right)
  +  \ll_2 \left( \ll_2
   \left(\begin{pmatrix}
   e_1 \\ \omega_1 \end{pmatrix}, \begin{pmatrix}
   e_2 \\ \omega_2 \end{pmatrix} \right), \begin{pmatrix}
   \xi \\ \rho \end{pmatrix} \right) \nonumber \\
   && =  \ll_3 \left( \begin{pmatrix} 0 \\ \diff \rho \end{pmatrix} ,
\begin{pmatrix} e_1\\ \omega_1 \end{pmatrix}, \begin{pmatrix} e_2\\ \omega_2 \end{pmatrix} \right) +
\ll_2 \left( \begin{pmatrix}
   \mathrm{L}_\xi e_1 + \ii [\rho, e_1] \\ \mathrm{L}_\xi \omega_1 + \ii [\rho, \omega_1] \end{pmatrix}, \begin{pmatrix}
   e_2 \\ \omega_2 \end{pmatrix} \right)    \nonumber \\
   && + \ll_2 \left( \begin{pmatrix}
   \mathrm{L}_\xi e_2 + \ii [\rho, e_2] \\ \mathrm{L}_\xi \omega_2 + \ii [\rho, \omega_2] \end{pmatrix}, \begin{pmatrix}
   e_1 \\ \omega_1 \end{pmatrix} \right)
  +  \ll_2 \left( \begin{pmatrix}
   - e_1 \wedge \diff \omega_2 - e_2 \wedge \diff \omega_1 - \diff \omega_1 \wedge e_2 - \diff \omega_2 \wedge e_1  \\ - \diff(e_1 \wedge e_2 + e_2 \wedge e_1) \end{pmatrix}, \begin{pmatrix}
   \xi \\ \rho \end{pmatrix} \right) \nonumber \\
   && = \ii \begin{pmatrix}  e_1 \wedge [\diff \rho, \omega_2] + e_2 \wedge [\diff \rho, \omega_1] + [\diff \rho, \omega_1] \wedge e_2 + [\diff \rho, \omega_2] \wedge e_1 \\
   [\diff \rho, [e_1, e_2]] \end{pmatrix} \nonumber \\
   &&\hskip -2cm+ \begin{pmatrix} - (\mathrm{L}_\xi e_1 + \ii [\rho, e_1]) \wedge \diff \omega_2 -  e_2 \wedge \diff (\mathrm{L}_\xi \omega_1 + \ii [\rho, \omega_1]) - \diff ( \mathrm{L}_\xi \omega_1 + \ii [\rho, \omega_1] ) \wedge e_2 - \diff \omega_2 \wedge (\mathrm{L}_{\xi} e_1 + \ii [\rho, e_1]) \\ -\diff ( (\mathrm{L}_{\xi} e_1 + \ii [\rho, e_1]) \wedge e_2 + e_2 \wedge (\mathrm{L}_{\xi} e_1 + \ii [\rho, e_1]) )  \end{pmatrix} \nonumber \\
   &&\hskip -2cm+ \begin{pmatrix} - (\mathrm{L}_\xi e_1 + \ii [\rho, e_1]) \wedge \diff \omega_2 -  e_2 \wedge \diff (\mathrm{L}_\xi \omega_1 + \ii [\rho, \omega_1]) - \diff ( \mathrm{L}_\xi \omega_1 + \ii [\rho, \omega_1] ) \wedge e_2 - \diff \omega_2 \wedge (\mathrm{L}_{\xi} e_1 + \ii [\rho, e_1]) \\ -\diff ( (\mathrm{L}_{\xi} e_1 + \ii [\rho, e_1]) \wedge e_2 + e_2 \wedge (\mathrm{L}_{\xi} e_1 + \ii [\rho, e_1]) )  \end{pmatrix} \nonumber \\
   && \hskip -2cm + \begin{pmatrix}
   \mathrm{L}_\xi ( e_1 \wedge \diff \omega_2 + e_2 \wedge \diff \omega_1 + \diff \omega_1 \wedge e_2 + \diff \omega_2 \wedge e_1 ) + \ii [\rho,  e_1 \wedge \diff \omega_2 + e_2 \wedge \diff \omega_1 + \diff \omega_1 + \diff \omega_1 \wedge e_2 + \diff \omega_2 \wedge e_1] \\
   + \mathrm{L}_\xi \diff ( e_1 \wedge e_2 + e_2 \wedge e_1 ) + \ii [\rho, \diff ( e_1 \wedge e_2 + e_2 \wedge e_1 )]
   \end{pmatrix} \nonumber\\
      &&= \begin{pmatrix} 0\\0 \end{pmatrix}\, . \label{homotopy3_2}
\end{eqnarray}
 The last equality in (\ref{homotopy3_2}) follows from the explicit cancellation of  all terms.
\\
\\
}

\bibliographystyle{ourstyle}
\bibliography{Reference.bib}

\end{document}